\documentclass[twocolumn,showpacs,preprintnumbers,amsmath,amssymb]{revtex4}
\usepackage{graphicx}
\input{epsf}

\newcommand{\sig}{\mbox{\boldmath{$\sigma$}}}
\newcommand{\ecal}{\mbox{\boldmath{$\cal E$}}}
\begin{document}

\title{Theory of resonant photon drag in monolayer graphene}
\author{M.V.Entin}
\affiliation{Institute of Semiconductor Physics,
Siberian Branch of Russian
Academy of Sciences, Novosibirsk, 630090, Russia}
\author{L.I.Magarill}
\affiliation{Institute of Semiconductor Physics,
Siberian Branch of Russian Academy of Sciences, Novosibirsk,
630090, Russia}
\author{D.L.Shepelyansky}
\affiliation{Laboratoire de Physique Th\'eorique (IRSAMC),
Universit\'e de Toulouse, UPS, F-31062 Toulouse, France}
\affiliation{LPT (IRSAMC), CNRS, F-31062 Toulouse, France}

\date{January  31, 2010 }


\begin{abstract}
Photon drag current in monolayer graphene with degenerate electron
gas is studied under interband excitation near the threshold of
fundamental transitions. Two main mechanisms  generate an emergence of electron current.
Non-resonant drag effect (NDE) results from
direct transfer of in-plane photon momentum ${\bf q}$ to electron and
dependence of matrix elements of transitions on ${\bf q}$.
Resonant drag effect (RDE)
originates from ${\bf q}$-dependent selection of transitions
due to a sharp form of  the Fermi distribution in energy.
The drag current essentially depends on the polarization
of radiation and, in general,  is not parallel to  ${\bf q}$.
The perpendicular current component
appears  if the in-plain electric field is tilted towards
 ${\bf q}$.
The RDE has no smallness connected with $q$ and exists in
a narrow  region of photon frequency $\omega$:
 $|\hbar\omega-2\epsilon_F|< \hbar sq$, where $s$ is the electron velocity.
\end{abstract}

\maketitle

\section{Introduction}
Though the theoretical study of two-dimensional carbon has a long
history \cite{wall},\cite{Slon},\cite{divinch},\cite{ando1} only
after experimental evidence of existence of graphene as a stable
two-dimensional crystal \cite{alpha},
\cite{berg,geim,zhang} this material became very
popular. The presence of zero gap and zero electron mass,
combined with a rather high mobility at room temperature,
makes graphene an unique material for various fundamental and applied
problems. At present  graphene is intensively studied both theoretically
and experimentally (see e.g. reviews
\cite{geim2,rev4}).

The study of graphene optics (see \cite{alpha3},\cite{falk}) is
stimulated by the prediction that the absorption in monolayer
graphene should be determined by the fundamental constant
$\alpha=e^2/\hbar c$ \cite{ando}, \cite{alpha2} and its
experimental evidence \cite{nair}. The investigation of coupling
between photons and electrons in graphene attracts now an active
interest of the community (see e.g. \cite{ziegler1,ziegler2}). An
observation of amplified stimulated terahertz emission from
optically pumped epitaxial graphene heterostructures has been
reported recently \cite{thzgener}. However, the photoinduced
currents, namely, photon drag and photogalvanic effects in
graphene were beyond of interest of the researchers. In this paper
we present the theoretical analysis of these effects.

The study  of light pressure on solids has rather  long history.
The simplest variant is an instantaneous transmission of photon
momentum to electrons. This process is permitted for interband
transitions or in presence of the ''third body'', for example,
phonons, other electrons, impurities. For a free particle this
process is forbidden by conservation laws. Small value of the
photon momentum makes Nonresonant photon Drag Effect (NDE) extremely weak.

At the same time there exists a less known variant of this effect,
namely Resonant photon Drag Effect (RDE) which has no weakness of
usual NDE
\cite{grinb},\cite{kast},\cite{alper}. Resonance drag occurs when
some partial kinetic property of electron gas sharply depends on
electron energy. A small photon momentum gives an
increase of the electron energy, that can drastically change the
relaxation time. This leads to  a significantly
different contributions to the electron current  for
electrons exited along or oppositely to the photon direction.
In \cite{alper} the situation was studied for interband transitions
in weakly doped GaAs when the electron energy approaches the
energy of longitudinal optical phonon. In this case electrons
exited along the direction of photon have larger energy than
electrons in opposite direction. Hence, their energy can exceed
the threshold for emission of optical phonon: they quickly emit
phonons and stop, while the opposite electrons will move freely
till they collide with impurity. This gives rise to the appearance of
charge flow in the direction opposite to the light ray.

Here we develop another idea for RDE based on a sharp Fermi
distribution which forbids the transitions below the Fermi energy
$\epsilon_F$. This idea is illustrated in Fig.1. Electrons are
excited from the hole cone to the electron cone  by photons with
frequency $\omega$ and wave vector ${\bf Q}$. The conditions for
resonant transitions are $sk+s|{\bf k-q}|=\omega$, $\hbar s|{\bf
k}|>\epsilon_F$, where ${\bf k}$ is the electron momentum counted
from the cone point, $s\approx 10^8$cm/s is the electron velocity,
and ${\bf q}$ is a projection of the wave vector ${\bf Q}$ of
radiation to the plane of graphene. The first condition determines
ellipse in ${\bf k}$ plane, the second limits a part of this
ellipse accessible for transitions. The wave vector tilts the
transitions towards its direction. Fig.~1 shows the case when the
frequency is close to $2\epsilon_F$. The electrons in the figure
are excited from the right segment of the Fermi surface contour.
This results in electron flow rightwards. Since $q\ll k_F$ the RDE
appears when the frequency is close to $2\epsilon_F$, namely if
$|\omega-2\epsilon_F|<sq$. Inside this window the current of RDE
has no smallness connected with $q$ and can be estimated as $j\sim
es\tau ~\pi\alpha~ P/(\hbar\omega)$, where $e$ and $s$ are the
electron charge  and the velocity, $\pi\alpha$ is the opacity of
graphene, $\tau$ is the transport relaxation time  and $P$ is the
light intensity. Physical meaning of this estimation is evident:
$\tau \pi\alpha P/(\hbar\omega)$ is the instantaneous density  of
exited electrons which conserve their momentum. Being multiplied
by the current of individual electrons $es$, this quantity gives
the current density.

Below we determine both NDE and RDE
for interband transitions in monolayer graphene with degenerate
electron gas.  Due to graphene electron-hole symmetry results are
applicable to n- and p-type graphene. In general the relaxation process for
electrons and holes are different that breaks
electron-hole symmetry. For concreteness, we consider the
n-type graphene. In this case the mean free time of excited electrons is
much longer than that of holes since due to different distance from
the Fermi level holes can easier emit phonons. Thus, the contribution of
holes will be neglected.

\begin{figure}
\includegraphics[width=8.5cm]{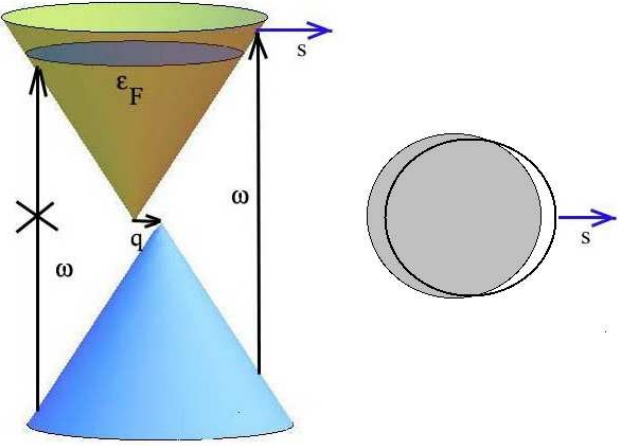}
\leavevmode \caption{(Color online) Interband phototransitions in
n-type graphene. Left panel: diagram of transitions in the
momentum-energy space. The hole cone is shifted in $\bf k$ space
by the photon wave vector $\bf q$. The transitions are permitted
only  above the Fermi level.  Right panel: projection to the
momentum plane. Filled circle represents the Fermi sea,
the elliptic
curve corresponds to the energy conservation equation
$s|{\bf k-q}|+sk=\omega$; only momenta outside the Fermi circle are
permitted corresponding to the right segment of the elliptic curve.
}
\end{figure}

Fig.~2 illustrates a possible experiment on excitation of the drag
current in a suspended graphene sheet placed in $(x,y)$ plane.
Light with frequency $\omega$, wave vector $\bf Q$  ($Q=\omega/c$)
and amplitude of electric field $\ecal$ illuminates graphene
plane. We consider transitions near the cone singularity. In this
case the current is determined by the projections of the electric
field  and the wave vector  onto the graphene layer.\footnote{The
vertical component of the electric field also interacts with
electrons, however, its action is weaker by the parameter $k_Fd$,
where  $d$ is the vertical distance between dangling bonds of
neighboring atoms. In fact, this component results in the
dynamical splitting of these states  and can be included in the
Hamiltonian as $\sigma_z eE_z d/2$. Comparison of this term with
considered one gives foregoing estimate.} These quantities are
${\bf E}\equiv{\bf e}E=({\cal E}_p\cos\beta,{\cal E}_s)$ and ${\bf
q}= (1,0)Q\sin\beta$, where $\beta$ is the  angle of incidence,
${\cal E}_s$ and ${\cal E}_p$ are amplitude components of the
electric field $\ecal$ perpendicular and parallel to the incident
plane. We ignore small modification of field caused by the layer.

\begin{figure}
\includegraphics[width=9cm]{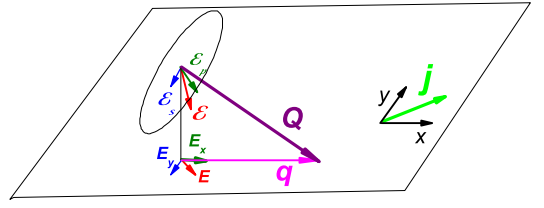}
\leavevmode \caption{(Color online)
Sketch of proposed experiment (see text for details).}
\end{figure}

\section{Basic equations}

The current of photon drag effect can be expressed via  the
probability of transition $g({\bf k})$  from
the hole state with a momentum ${\bf k-q}$ to the electron state with momentum ${\bf k}$
and the electron velocity ${\bf v}({\bf k})=s{\bf k}/k$ as
\begin{equation}
\label{j}
    {\bf j}=4e\int \frac{d{\bf k}}{4\pi^2} {\bf v}({\bf k})\tau g({\bf k}),
\end{equation}
where the coefficient 4 accounts for the valley and spin
degeneracies.  The dependence on the photon momentum results from
the momentum and energy conservation laws and the matrix elements
for transition. For simplicity we put below $\hbar=1$.

The two-band Hamiltonian near the Dirac point is
\begin{eqnarray}
\label{1}
   \hat{H}({\bf k})=s\left(
\begin{array}{cc}
  0 & k_x-ik_y \\
  k_x+ik_y&0\\
\end{array}
\right)=s{\bf k}\sig.
\end{eqnarray}
Here $\sig$ is the vector of the Pauli matrices.
The eigenvalues and eigenvectors of the Hamiltonian (\ref{1}) are
$\epsilon_{\pm}({\bf k})=\pm sk$ and
$\Psi_{\pm}({\bf k})=(1,\pm e^{i\phi_{\bf k}})/\sqrt{2}$,
where $\phi_{\bf k}$ is the polar angle of the vector ${\bf k}$.
The different signs correspond to
electrons and holes. The interaction with the wave is determined
by the matrix elements of the velocity $\nabla_{\bf k}
\hat{H}({\bf k})=s\sig$ between the hole and electron states with the
momenta ${\bf k-q}$ and
${\bf k}$: ${\bf v}^{-+}=(\Psi_-({\bf k-q})^*s\sig\Psi_+({\bf k}))$, correspondingly.

 The transition probability $g({\bf k})$ is
\begin{equation}
\label{g} g({\bf k})=\frac{\pi e^2}{2\omega^2}|{\bf E}{\bf
v}^{+-}|^2\delta(sk+s|{\bf k}-{\bf q}|-\omega)\theta(\epsilon_{\bf
k}-\epsilon_F),
\end{equation}
where $\theta(t)$ is the Heaviside function.
The expression for current Eq.(\ref{j}) can be rewritten as
\begin{equation}\label{jj}
    {\bf j}=\frac{e^3E^2s^3}{2\pi\omega^2}\int d{\bf k} \tau   \frac{\bf k}{|{\bf k}|}
   a_{jk}e_je_k\delta(sk+s|{\bf k-q}|-\omega)\theta(sk-\epsilon_F)\end{equation}
    where
    \begin{equation}\label{a}a_{jk}=
    \frac{1}{s^2}v_j^{-+*}v_k^{-+}.
\end{equation}
We utilized the symmetry of the tensor $a_{ij}$ resulting to inclusion of the
field polarization in the combinations $e_i^*e_j+e_j^*e_i$ only
and independence on the degree of circular polarization. Hence,
without loss of generality one can consider the field as
linear-polarized and ${\bf e}$ as real.

Due to  the smallness of the wave vector $q$,
as compared to the electron momentum, one can expand all quantities in
powers of  $q$.
Expanding by $q$ we can write the argument of the delta-function
as $sk+s|{\bf k}-{\bf q}|-\omega\approx 2sk-\omega -sq \cos\phi_{\bf k}$
(we choose the direction of axis x along ${\bf q}$).
At the same time, $q$ is  comparable with  $2sk-\omega$
and we keep ourselves from subsequent expansion of the delta-function.

Expanding the tensor $a_{ij}$, we have
\begin{eqnarray}
\label{aa}
a_{xx}&=&\sin^2\phi_{\bf k}(1+\frac{q}{k}\cos\phi_{\bf k}),\nonumber\\
a_{yy}&=&\cos^2\phi_{\bf k}-\frac{q}{k}\sin^2\phi_{\bf k}\cos\phi_{\bf k},\\
2\mbox{Re}(a_{xy})&=&-2\sin\phi_{\bf k}\cos\phi_{\bf k}-\frac{q}{k}\sin\phi_{\bf k}\cos(2\phi_{\bf k})\nonumber
\end{eqnarray}
From Eq.(\ref{jj}) we obtain for components of the current
\begin{eqnarray} \label{jjj}
j_x=-2J_0\int_{-1}^{min(1,a)}
\frac{dx}{\sqrt{1-x^2}}\frac{\tau}{\tau_0}\times\nonumber~~~~~~~~~~~~~~~~~\\
\Big\{e_x^2 [-x(1-x^2)(1+bx)+2b(1-x^2)(1-2x^2)]+ \nonumber \\
 e_y^2[-x^3(1+bx)+4bx^2(1-x^2)]\Big\} ;\\
j_y=2J_0e_xe_y\int_{-1}^{min(1,a)} dx\frac{\tau}{\tau_0}\sqrt{1-x^2}\times\nonumber~~~~~~~~~\\
 \Big\{-2x(1-x^2)(1+bx)+2b(3x^2-1)\Big\}.
\end{eqnarray} Here we have introduced the following notations:
$$J_0=\frac{e^2}{\hbar c}\frac{cE^2}{8\pi \hbar\omega}|e|\tau_0s, ~~\tau_0=\tau|_{k=k_F} ,~~   a=\frac{\omega-2\epsilon_F}{sq}, ~~ b=\frac{sq}{\omega}.$$

If $\tau$ is independent on the energy of electrons then the integration in
Eq.(\ref{jjj}) can be done directly.
The current has different values inside and outside the region
$|\omega-2\epsilon_F|<sq$. If $|\omega-2\epsilon_F|<sq$ then we have
\begin{eqnarray}
\label{RDE}
j_x&=&-\frac{2}{3}J_0\sqrt{1-a^2}((1-a^2)e_x^2+(2+a^2)e_y^2),\\
j_y&=&-\frac{4}{3}J_0(1-a^2)^{3/2}e_xe_y.
 \end{eqnarray}
These values represent resonant photon drag RDE. It remains constant if
$q\to 0$. The value of resonant current is determined by $J_0$.
For the photon flow
$cE^2/8\pi\hbar\omega=10^{19}$cm${}^{-2}$s${}^{-1}$,
$\tau=10^{-12}$ s, $J_0=1.16\cdot 10^{-6}$A/cm. This approximately
corresponds to a power of $0.1W/cm^2$ for photons with energy $0.1eV$.

If $|\omega-2\epsilon_F|>sq$, then there is only NDE current. It is proportional to $q$:
\begin{eqnarray} \label{NDE}
j_x&=&J_0\frac{\pi s q}{4\omega}(3e_x^2-e_y^2),\\
j_y&=&\frac{3}{2}J_0\frac{\pi s q}{\omega}e_xe_y.
 \end{eqnarray}
The value of NDE is significantly smaller then the RDE value.

In agreement with the simple estimates the RDE has always the direction
opposite to the direction of light wave vector. Its polarization
dependence is explained by the dependence of the directional
diagram of excitation: most of carriers are excited perpendicular
to the polarization. At the same time the Fermi sea limits the
transitions by the direction of the photon wave vector. This
circumstances together determine lower x-component of current  if
${\bf e}||{\bf q}$ in comparison with the case ${\bf e}\perp{\bf
q}$ and also the appearance of  y-component of the RDE current.

In agreement with the system symmetry, $j_y$ exists only if the
polarization has both $e_x$ and $e_y$ components. The RDE current
exists in a narrow window $|\omega-2\epsilon_F|<sq$ which shrinks
if $q\to 0$. But inside this window RDE is much stronger than NDE
so the later can be neglected in this window.
\begin{figure}
\includegraphics[width=7.5cm]{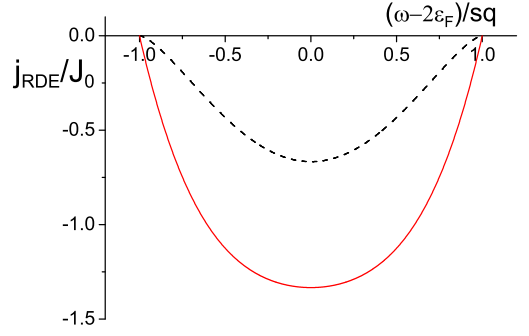}
\leavevmode \caption{ Resonant photon drag current in units of
$J_0$ versus normalized frequency $(\omega-2\epsilon_F)/sq$. The
solid curve shows the longitudinal component of  current $j_x$,
the field is polarized along the projection of the wave vector on
the plane ($\theta=0$) and $j_y$ at $\theta=\pi/4$.  The dashed
curve shows  $j_x$ at $\theta=\pi/2$. }
\end{figure}
\begin{figure}[h]
\includegraphics[width=7.5cm]{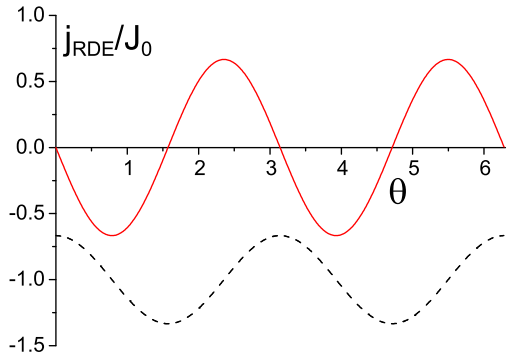}
\leavevmode \caption{Polarization dependence of the RDE current at
$\omega=2\epsilon_F$;  $j_x$ is shown by  dash curve, $j_y$ is
shown by solid curve.}
\end{figure}

The sign of x-component of NDE depends on polarization. This
contradicts to a simple assumption according to which the current
is mainly determined by kicks which photons give to electrons.
The origin of
this difference is the dependence of the directional diagram on the
small wave vector ${\bf q}$ via the parameter $a_{ij}$: at some
polarizations electrons prefer to be excited in opposite direction
to ${\bf q}$. This explains the change of sign.

Fig.~3 demonstrates the dependence of RDE current components on
the frequency in the window  $|\omega-2\epsilon_F|<sq$ where RDE
exists. The current vanishes at the edges of the window. The
component $j_x$ is larger for the polarization along the y axis.
The component $j_y$ appears only for tilted polarization of the
light. Fig.~4 shows the dependence of $j_x$ and $j_y$ on the angle
$\theta$ between the vector of polarization ${\bf e}$  and the
wave vector ${\bf q}$.

\section{Discussion}
We have studied the electron contribution to the photon drag current. In
fact, in the considered system the hole contribution also
presents. The symmetry between holes and electrons in a neutral
system means that these contributions double. However, the result
will be changed if to take into account the difference between
electrons and holes caused by their different excitation energy:
while electrons are generated near the Fermi energy the holes
appear well below the Fermi energy. This leads to a strong
difference between the relaxation times. In high-mobility samples
at low temperature  the momentum relaxation time near the Fermi
energy is much greater than far from the Fermi energy. At the same
time, quick relaxation of excited electrons (holes) to the Fermi
energy due to electron-electron interaction (described by e-e
relaxation time $\tau_{ee}$) conserves their momenta up to the
moment when excitations reaches the temperature layer. This
results in equality of holes and electrons contributions to the
current. And vice versa, electron-phonon relaxation  can cancel
the hole contribution if $\tau_{e-ph} \ll\tau_{ee}$, where
$\tau_{e-ph}$ is the time of energy relaxation due to
electron-phonon collisions. Thus, the obtained current should be
multiplied by a factor 2 in the case of quick e-e relaxation and
be kept unchanged in the opposite case. We note, that when the Fermi
energy tends to zero the system becomes symmetric.

The RDE exists in a narrow energy range
$\Delta \epsilon \approx \hbar s q \approx \hbar \omega  s/c$
near the Fermi energy. This means that the RDE is visible for
temperature $T< \Delta \epsilon$. For photons with $\hbar \omega =0.1 eV $
this gives $T< 3 K$.

The  observation of the resonant photon drag in monolayer graphene
is accessible to the modern experimental technique  that allows to
investigate interesting aspects of coupling between photons and
electrons in this material.

\section{Acknowledgments}
We thank A.D.Chepelianskii for useful discussions.
The work was supported by grant of RFBR No 08-02-00506 and No
08-02-00152 and ANR France PNANO grant NANOTERRA;
MVE and LIM thank Laboratoire de Physique Th\'eorique, CNRS
for hospitality during the period of this work.


\begin{thebibliography}{99}
\bibitem{wall}P.R.~Wallace, Phys. Rev. {\bf 71}, (1947) 622;
\bibitem{Slon} J.C.~Slonczewski and P.R.~Weiss, Phys. Rev. {\bf 109}, (1958) 272.
\bibitem{divinch} D.P.~DiVincenzo and E.J.~Mele, Phys. Rev. B {\bf 29}, (1984) 1685.
\bibitem{ando1} T.~Ando, T.~Nakanishi, and R.~Saito, J. Phys. Soc. Japan
               {\bf 67}, (1998) 2857.
\bibitem{alpha}K.S.~Novoselov, A.K.~Geim, S.V.~Morozov, D.~Jiang, Y.~Zhang, S.V.~Dubonos,
            I.V.~Grigorieva, and A.A.~Firsov, Science, {\bf 306}, 666 (2004).
\bibitem{berg} C.~Berger, Z.~Song, X.~Li, X.~Wu, N.~Brown, C.~Naud, D.~Mayou, T.~Li,
           J.~Hass, A.N.~Marchenkov, E.H.~Conrad, P.N.~First, and W.A. de Heer,
           Science {\bf 312}, 1191 (2006).
\bibitem{geim} K. S. Novoselov, A.K.~Geim, S.V.~Morozov, D.~Jiang, M.I.~Katsnelson,
            I.V.~Grigorieva, S.V.~Dubonos, A.A.~Firsov , Nature {\bf 438}, 197 (2005).
\bibitem{zhang} Y.~Zhang, J.W.~Tan, H.L.~Stormer, and P.~Kim, Nature
                {\bf 438}, 201 (2005).
\bibitem{geim2}A.K.~Geim and K.S.~Novoselov, Nature Materials {\bf 6}, 183
              (2007).
\bibitem{rev4} A.H.~Castro Neto, F.~Guinea, N.M.R.~Peres, K.S.~Novoselov,
           and A.K.~Geim,  Rev. Mod. Phys.,  {\bf 81}, 109 (2009).
\bibitem{alpha3} L.A.~Falkovsky and A.A.~Varlamov, Eur. Phys. J. B,
         {\bf 56}, 281 (2007).
\bibitem{falk} L.A.~Falkovsky,   Phys. Usp. {\bf 51}, 887 (2008)
            [Usp. Fiz. Nauk,{\bf 178}, 923 (2008)]
\bibitem{ando} T.~Ando, Y.~Zheng and H.~Suzuura, J. Phys. Soc. Japan, {\bf 71},
              1318 (2002).
\bibitem{alpha2} V.P.~Gusynin, S.G.~Sharapov and J.P.~Carbotte, Phys.
            Rev. Lett. {\bf 96}, 256802 (2006).
\bibitem{nair} R.R.Nair, P.Blake, A.N.Grigorenko, K.S.Novoselov, T.J.Booth,
          T.Stauber, N.M.R.Peres, A. K. Geim, Science, {\bf 320},
         1308 (2008).
\bibitem{ziegler1} J.Z.~Bern\'ad, U.Z\"ulicke, and K.~Ziegler,
         arXiv:1001.3239[cond-mat] (2010).
\bibitem{ziegler2} K.~Ziegler and A.~Sinner, arXiv:1001.3366[cond-mat] (2010).
\bibitem{thzgener} T.Otsuji, H.Karasawa, T.Komori, T.Watanabe, H.Fukidome,
        M.Suemitsu, A.Satou,  and Victor Ryzhii,
        arXiv:1001.5075[cond-mat] (2010).
\bibitem{grinb} A. A. Grinberg, Zh. Eksp. Teor. Fiz. {\bf 58}, 989 (1970) [Sov.
           Phys.-JETP {\bf 31}, 531 (1970)].
\bibitem{kast} A.M.Danishevskii, A.A.Kastal'skii, S.M.Ryvkin, and
          I.D.Yaroshetskii, Zh.Exp.Teor.Fiz. {\bf 58}, 544 (1970)
         [Sov. Phys. JETP, {\bf 31}, 292 (1970)].
\bibitem{alper} V.L.~Al'perovich, V.I.~Belinicher, V.N.~Novikov, and A.S.~Terekhov,
          {\bf 33}, 557 (1981)
         [Sov. Phys.-JETP Lett.,  {\bf 33}, 573 (1981)].
\end{thebibliography}
\end{document}